# GNSS Reflectometry and Remote Sensing: New Objectives and Results

Shuanggen Jin[1] and Attila Komjathy[2]

[1]Center for Space Research, University of Texas at Austin, TX 78759, USA
[2]Jet Propulsion Laboratory, NASA, Pasadena, CA 91109, USA

Email: sgjin@csr.utexas.edu; shuanggen.jin@gmail.com
Tel: 512-795-7669; Fax: 512-471-3570

**Abstract**

The Global Navigation Satellite System (GNSS) has been a very powerful and important contributor to all scientific questions related to precise positioning on Earth's surface, particularly as a mature technique in geodesy and geosciences. With the development of GNSS as a satellite microwave (L-band) technique, more and wider applications and new potentials are explored and utilized. The versatile and available GNSS signals can image the Earth's surface environments as a new, highly precise, continuous, all-weather and near-real-time remote sensing tool. The refracted signals from GNSS Radio Occultation satellites together with ground GNSS observations can provide the high-resolution tropospheric water vapor, temperature and pressure, tropopause parameters and ionospheric total electron content (TEC) and electron density profile as well. The GNSS reflected signals from the ocean and land surface could determine the ocean height, wind speed and wind direction of ocean surface, soil moisture, ice and snow thickness. In this paper, GNSS remote sensing applications in the atmosphere, oceans, land and hydrology are presented as well as new objectives and results discussed.

## 1. Introduction

The Global Navigation Satellite System (GNSS), including the Global Positioning System (GPS) in the United States, the Russian GLONASS, the European Galileo and the Chinese COMPASS (Beidou), can be characterized as a highly precise, continuous, all-weather and near-real-time microwave (L-band) technique with signals through the Earth's atmosphere. These characteristics of GNSS imply more and wider applications and potentials. When the GPS signal propagates through the Earth's atmosphere, it is delayed by the atmospheric refractivity, which results in lengthening of the geometric path of the ray. In 1992 when the GPS became fully operational, Ware (1992) suggested limb sounding the Earth atmosphere using GPS atmospheric delay signals. On 3 April 1995, the small research satellite of Microlab-1 was successfully put into a Low Earth Orbit (LEO) to validate the GPS radio occultation method (Feng and Herman, 1999). Since then, the GPS/Meteorology Mission (GPS/MET) using the radio occultation technique has been used to produced accurate, all weather, global refractive index, pressure, density profiles in the troposphere, temperature with up to the lower stratosphere (35-40 km), and the ionospheric total electron content (TEC) as well as electron density profiles (Rocken, 1997; Hajj and Romans, 1998; Syndergaard, 2000; Derek and Benjamin, 1999), to improve weather analysis and forecasting, monitor climate change, and monitor ionospheric events.

In addition, surface multi-path is one of main error sources for GNSS navigation and positioning. It has recently been recognized, however, that the special kind of GNSS multi-path delay reflected from the Earth's surface, could be used to sense the Earth's surface environments. Hall and Cordey (1988) first addressed the Bistatic radar using L-band signals transmitted by GPS proposed by the European Space Agency (ESA) as an ocean scatterometer. Rubashkin (1993) demonstrated the concept of bistatic radar sensing of the ocean surface using two satellites with a transmitter in a low Earth orbit and a receiver in geosynchronous orbit. Martı́n-Neira (1993) proposed and described an altimeter system using ocean GPS reflections to measure sea surface heights. Katzberg and Garrison (1996) proposed the reflection of the GPS signal from the ocean for ionospheric measurements by adding a GPS receiver and downward-pointing antenna to any satellites, and evaluated the feasibility and effectiveness. Later a number of experiments and missions using GPS reflected signals from the ocean and land surface have been tested and applied, such as determining ocean surface height, wind speed and wind direction of ocean surface, soil moisture, snow and ice thickness (Komjathy, et al., 1999; Rius et al., 2002; Wagner and Kloko, 2003; Germain et al., 2004; Komjathy et al., 2004; You et al., 2004; Kostelecký et al., 2005; Thompson et al., 2005; Gleason et al., 2005).

Therefore, the versatility and availability of GNSS reflected and refracted signals result in many new applications. The sensitivity of these signals to propagation effects is useful for various environmental remote sensing. This paper will address new objectives and results of GNSS remote sensing in the atmosphere, oceans, land and hydrology as well as new opportunities for future missions.

## 2. GNSS atmospheric remote sensing

Due to the atmospheric refraction, GPS signals propagate through the Earth atmosphere along a slightly curved path and with slightly retarded speeds (Figure 1a). For a long time, the delay of GNSS signals in the troposphere and ionosphere was considered as a nuisance, an error source, and now it has been used to determine the useful atmospheric parameters, including tropospheric water vapor, temperature and pressure, and ionospheric total electron content (TEC) and electron density profile (e.g. Jin et al. 2006). Nowadays, a number of GPS Radio Occultation (RO) missions have been successfully launched for atmospheric and ionospheric detections and climate change related studies, such as the US/ Argentina SAC-C, German CHAMP (CHAllenging Minisatellite Payload), US/Germany GRACE (Gravity Recovery and Climate Experiment), Taiwan/US FORMOSAT-3/COSMIC (FORMOsa SATellite mission - 3/Constellation Observing System for Meteorology, Ionosphere and Climate) satellites, the German TerraSAR-X satellites and the European MetOp. These GPS RO satellites together with ground-based GPS observations have provided the tropospheric water vapor, pressure, temperature, tropopause parameters, ionospheric TEC and electron density profiles, which were consistent with traditional instruments observations at comparable accuracies (e.g. Schmidt et al., 2005; Jin et al., 2006 and 2007; Schmidt, et al. 2008). For example, Figure 2 shows ionospheric electron density profiles from ground-based GPS tomography reconstruction over South Korea on 28

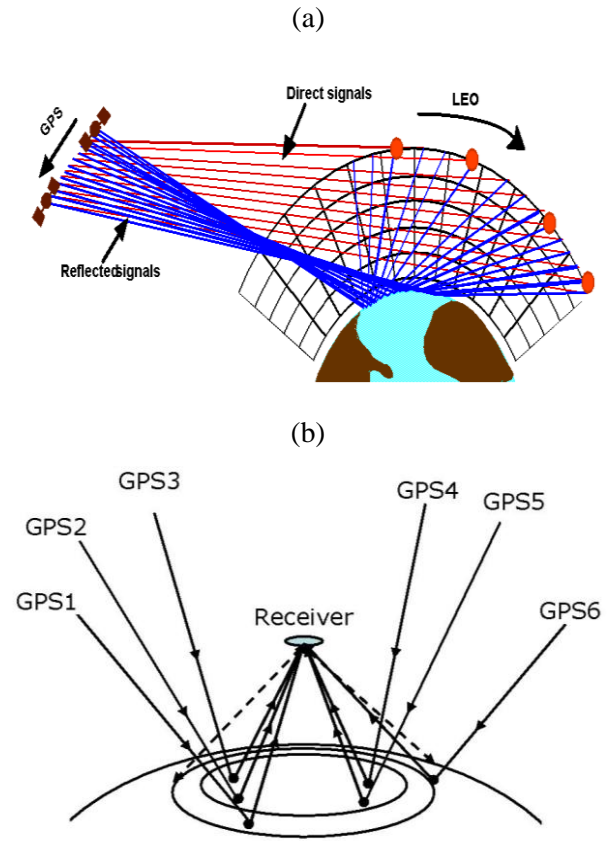

Figure 1. GNSS refracted and reflected signals and geometry (Yunck, 2003).

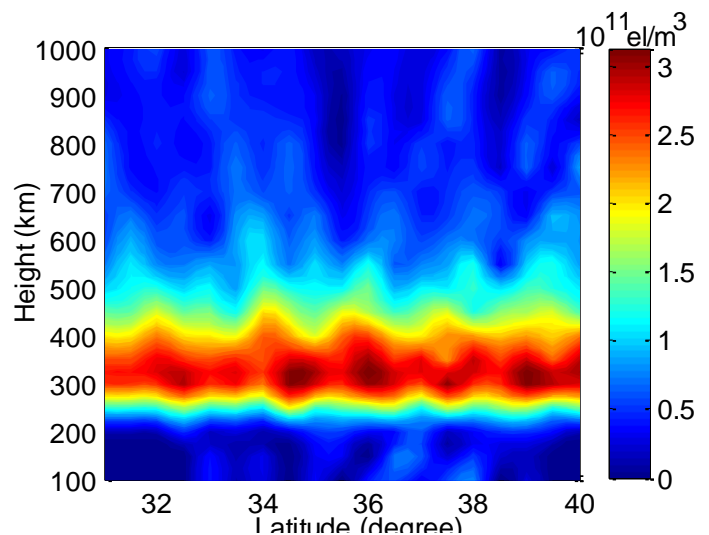

Figure 2. Ionospheric electron density profiles from ground-based GPS tomography reconstruction in South Korea on 28 October 2003 at UT: 13:00.

October 2003 at UT 13:00 (Jin et al. 2006). These ground-based and space-borne GPS observations can provide important 3-D ionospheric profile information related to various activities and states in the ionosphere, particularly for solar flares and geomagnetic storms (e.g., Jakowski et al. 2007; Jin et al. 2008).

The traditional atmospheric observing instruments, such as the water vapor radiometer (WVR), ionosonde, incoherent scatter radars (ISR), topside sounders onboard satellites, in situ rocket and satellite observations (e.g. Cloud-Aerosol Lidar and Infrared Pathfinder Satellite Observation (CALIPSO)), are expensive and also partly restricted to either the bottomside ionosphere or the lower part of the topside ionosphere. While GPS satellites in high altitude orbits (~20,200 km) are capable of providing details on the structure of the entire ionosphere, even the plasmasphere. Therefore, GPS has been widely applied in atmospheric sounding, meteorology, climatology and space weather. Furthermore, GPS can monitor the ionospheric disturbances during earthquakes and volcanoes due to the coupling of the solid-Earth and the ionosphere, e.g. the 2008 Wenchuan Earthquake in China (Jin et al., 2010), which may provide some new insights on solid-Earth activities.

**3. GNSS ocean remote sensing**

The GPS satellites are constantly broadcasting radio signals to the Earth. However, part of the signals is reflected back from the rough Earth's surface (Figure 1b). The delay of the GPS reflected signal with respect to the rough surface could provide information on the differential paths between direct and reflected signals. Together with information on the receiver antenna position and the medium, the delay measurements associated with the properties of the reflecting surface can be used to produce the surface roughness parameters and to determine surface characteristics. For example, the measurements of GPS reflected signals from the ocean surface could retrieve the ocean surface height, wind speed, wind direction, and even sea ice conditions.

**3.1 Determining ocean surface height**

Conventional single radar remote sensing of the oceans requires dedicated transmitters and receivers with large directional antennae top to achieve high resolution, while GPS as a bistatic radar system, requires only a receiver, not a transmitter, and can receive global, continuous, near-real-time and all weather signals from global covering GPS satellites. Although the satellite radar technique measures sea surface height at high spatial resolution along its ground track, the cross track distance is usually quite large (e.g., about the order of 300 km for TOPEX/Poseidon). Furthermore, the temporal resolution is lower with the 10-day repeat orbit of TOPEX/Poseidon, so it is too coarse for monitoring the sea surface. The new wide-swath altimeter (Le Hénaff et al., 2008) has improved the spatial coverage, however, the temporal resolution is still 10 days. The GPS reflected signals from the ocean surface can measure the ocean surface height with high temporal-spatial resolution. The tested results in the determining the sea surface height using GPS reflected signals showed a good consistency at the level of about 2 cm in very calm sea state with independent TOPEX/Poseidon data (Martin-Neira et al., 2001). Therefore, the new applications of GPS reflected signals from the ocean surface could compensate these defects of existing techniques, e.g., radar altimetry.

**3.2 Monitoring the ocean surface winds**

GPS technique in a bistatic radar configuration also can measure the wind vector parameters on the sea surface. The key issue is to extract information from the GPS reflected signal. The primary measurement is the received power from the GPS reflected signal for a variety of delays and Doppler values in a glistening zone surrounding a nominal specular reflection point (Garrison et al., 1997; Clifford et al., 1998). The size and shape of the glistening zone are functions of the roughness of the ocean surface. As the receiver-generated pseudorandom noise codes are delayed in time with respect to directly received line-of-sight signals, the reflected power from the glistening zone are measured. The shape of the resulting waveform of power-versus-delay is dependent on the roughness of the ocean surface related to the surface wind speed and direction parameters. Therefore, the GPS reflected signal power measurements can measure the ocean surface wind speed and direction (e.g. Armatys et al., 2000; Lin et al., 1998; Komjathy et al., 2000a; Garrison et al., 2002; Cardellach et al., 2003; Thompson et al., 2005).

Using the Delay Mapping Receiver (DMR) with the L1 GPS signal at 1575.42 MHz developed at the NASA-Langley Research Center, scientists from the NASA Langley Research Center and later, Purdue University and University of Colorado at Boulder have successfully estimated the wind speeds and directions on the ocean surface with high accuracy. The estimated wind speed using surface-reflected GPS data is consistent with independent wind speed measurements derived from the TOPEX/Poseidon altimetry satellite and balloon measurements at the level of 2 m/s. The estimated wind direction agrees with results obtained from buoys (Zavorotny and Voronovich, 2000; Garrison, 1999; Komjathy et al., 2001; Cardellach et al. 2003). In addition, the surface wind speed retrievals from GPS reflected signals have been shown to be consistent with other independent techniques such as scatterometers, flight level winds, microwave radiometers, and dropsondes (Katzberg, et al. 2006a, and 2009). Furthermore, the reflected GPS signals from the United Kingdoms Low Earth Orbit DMC (Disaster Monitoring Constellation) satellite also showed successful results with low wind speeds and limited accuracy using this technique in ocean remote sensing (Gleason et al. 2005).

### 3.3 Sounding the sea ice conditions

Due to complex and varying conditions of sea ice, e.g. an inaccessible environment and persistent cloud covering, it is very difficult to monitor sea ice conditions with conventional instruments. Thus, measuring sea ice conditions mainly relies on satellite radar techniques. However, no single sensor is capable of providing the essential range of observations (Livingstone et al., 1987; Rubashkin et al., 1993). For example, Synthetic-aperture radar (SAR) images have sufficient spatial resolution to resolve detailed ice features, but repeat times of existing satellites are relatively long when compared to the change rate of open water fraction in the ice pack, although this aspect may be improved with more satellites in the future. Furthermore, SAR data carry a substantial penalty in cost for image acquisition and processing. In addition, space-borne passive microwave sensors may provide more frequent coverage at several wavelengths, but they have substantially lower spatial resolution. While optical and thermal sensors provide a middle ground in resolution and temporal sampling between SAR and passive microwave satellites, they are limited by cloud cover and visibility conditions.

The GPS reflectometry can measure the sea ice conditions as a new technique. Komjathy et al. (2000b) analyzed the aircraft experiment of GPS reflections from Arctic sea ice and over the ice pack near Barrow, Alaska, USA. Correlations from comparisons between RADARSAT backscatter and GPS forward scattered data indicate that the GPS reflected signals could provide useful information on sea ice conditions. The reflected signal shape was quite consistent for the moderate altitudes of the airborne GPS receiver and the peak power changed significantly along the flight track. This behavior of the reflected signal showed clearly the sensitivity to ice condition, indicating that the GPS reflected signals can be used to determine the ice features. In addition, as the effective dielectric constant of ice depends on various factors, such as the ice composition, density, age, origin, salinity, temperature, morphology (Shohr, 1998), the internal ice states can be determined with the reflection coefficient over a frozen sea surface by the effective dielectric constant of ice and the dielectric constant of the underlying water under some conditions (Melling, 1998). In the future, GPS reflected signals might provide more detailed information and internal states of sea ice, including the floe ridges, frost flowers, broken ice, and fine-scale roughness at the snow-ice interface. Therefore, the reflected GPS signals have a high potential and applications in sensing and investigating the sea ice state, particularly for inaccessible and atrocious sea ice cover.

## 4 GNSS land/hydrology remote sensing

### 4.1 Monitoring lake and wetland

For the conventional geodetic instruments, it is difficult to monitor the lake. While cross-track distance of satellite radar altimetry is usually large, the spatial resolution is too low to monitor small regions. GPS reflected signals could detect characteristics of small Earth's surface, such as the lake. Treuhaft et al. (2001) measured the surface height of the Crater Lake in Central Oregon with 2-cm precision in 1 second using the GPS signals reflected from the surface of Crater Lake. In addition, the strength of the reflected signal is also a discriminator between wet and dry ground areas and therefore could be applied to coastal and wetland mapping (Garrison and Katzberg, 1998)

### 4.2 Measuring soil moisture

The soil moisture content is important for hydrology, climatology, and agriculture (Jackson et al., 1996). Measuring the soil moisture content may predict potential flood hazards, understand land-atmosphere energy balance, and expect crop yield. In the past, the soil moisture was inferred using passive radiometers and active radar sensors (e.g., Njoku and Entekhabi, 1996; Ulaby et al., 1996; Mancini et al., 1996; Jackson 2001; Jackson et al., 2002). However, these two remote sensing techniques are more expensive. The GPS signal forward scattered from a land surface is similar to scattering from an ocean surface (Garrison and Katzberg, 1998), while the main differences are in the spatially and temporally varying dielectric constant, surface roughness, and possible vegetative cover. Masters et al., (2000; 2004) obtained soil moisture at a USDA/SCAN (United States Department of Agriculture/Soil Climate Analysis Network) site, located on the Central Plains Experimental Range of Colombia, with the peak power of the GPS reflections (Figure 3). Katzberg et al (2006b) has also successfully estimated soil reflectivity and dielectric constant with GPS reflected signals.

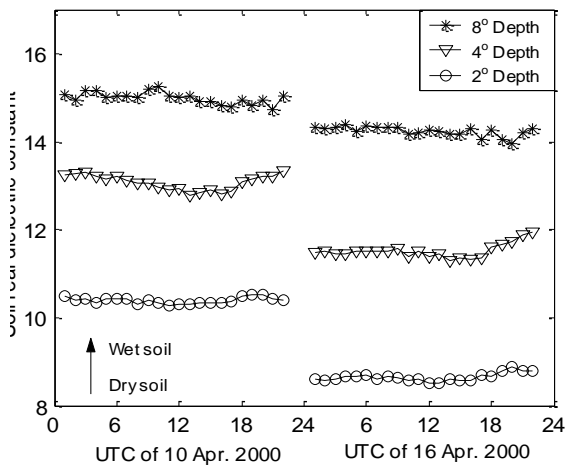

Figure 3. Soil moisture dielectric constant at the USDA/SCAN (United States Department of Agriculture/Soil Climate Analysis Network) site on the Central Plains Experimental Range of Colombia from peak power of the GPS reflections, modified from Masters et al. (2000).

In addition, the multi-path from ground GPS network may be related to the near surface soil moisture. A recent interesting result on fluctuations in near surface soil moisture from a 300 m$^2$ area has been successfully retrieved from the ground GPS multi-path (Larson et al., 2008), fairly matching soil moisture fluctuations in the top 5 cm of soil measured with conventional sensors. The results indicate a potential of remotely sensing soil moisture content using ground GPS multi-path signals. In the future, the existing global GPS networks with more than thousands of GPS receivers operated around the world may provide a tool to estimate global land soil moisture in near real-time for hydrology and climate studies, particularly the continuous IGS (International GNSS Service) network.

### 4.3 Measuring land snow/ice thickness

Snow and ice on the land are important components of climate systems and a critical storage component in the hydrologic cycle as well. However, in situ observations of snow distribution are sparse, and remotely sensed products are imprecise and only available at a coarse spatial scale, e.g., the U.S. Snowpack Telemetry (SNOTEL) network (Serreze et al., 1999). As the ice thickness is related to the amplitude of the reflected signal as a function of the incidence angle or relative amplitudes between different polarizations (Lowe et al., 2002), the snow/ice thickness can be retrieved from the GPS reflected signals. In addition, the change in snow depth is also monitored using the corresponding multi-path modulation of the ground GPS signal. The tested results for two spring 2009 snowstorms in Colorado showed strong agreement between GPS snow depth estimates, field measurements, and nearby ultrasonic snow depth sensors (Larson et al., 2009).

### 5. Conclusion and discussion

The Earth's surface environments are traditionally measured using the radar techniques with the dedicated transmitters and receivers at generally lower temporal-spatial resolution, e.g. about the 10-day repeat orbit and the order of 300 km for TOPEX/Poseidon. While the GPS with only a receiver, not a transmitter can measure the Earth's surface environmental parameters as a new, highly precise, continuous, all-weather and near-real-time remote sensing tool. The refracted signals from GNSS Radio Occultation satellites together with global ground GNSS observations can produce high-resolution atmospheric parameters, e.g. the tropospheric water vapor, temperature and pressure, tropopause parameters and ionospheric TEC and electron density profile. These atmospheric parameters are very useful in meteorology, climatology and space weather as well as understanding the coupling of the solid-Earth and the atmosphere. The GNSS reflected signals from the ocean and land surface can determine the ocean height, wind speed and wind direction of ocean surface, soil moisture, ice and snow thickness, which are consistent with other independent techniques such as radar altimetry, buoys, scatterometers, flight level winds, microwave radiometers, and dropsondes. Therefore, the surface reflected and atmospheric refracted GPS signals, for L-band observations and simple devices useable on any type of aircraft, are expected to revolutionize various atmospheric sounding, ocean remote sensing and land/hydrology mapping. Once the delay-Doppler-mapping GPS receiver is installed onboard a satellite or spacecraft, it will provide us with the unique opportunity to use GPS remote sensing tool to infer various environmental parameters.

With more GNSS satellite constellations in the future, the reflected and refracted GPS signals will soon become a more powerful source of data for scientists to get better understanding of global ocean/land surface characteristics, ocean safety, climate change and global warming, especially for various geohazards, complex topographic land and larger ocean in the world. Also the new developments and applications are expected for the future space-based missions at a high temporal-spatial resolution, e.g. Shuttle measurements. Furthermore, it should be free for public community and cover the entire globe as well. In addition, the bistatically reflected signals from the GNSS needs to be further extended to monitor the more details of Earth's surface characteristics in the future, such as the ocean wave and eddy, ocean salinity, internal states of sea ice, storm surge height, tsunami wave propagation and dynamics, land and flow motions, landslide and subsidence as well as geoharzard warning system.